# Performance Improvement of Dimmable OFDM-Visible Light Communication using Subcarrier Index Modulation and Reed Solomon Encoding

Nima Taherkhani[a], Kamran Kiasaleh[a*]

*Abstract*—**In this paper, we propose a new subcarrier index modulation scheme for orthogonal frequency division multiplexing (OFDM), which incorporates the Reed-Solomon (RS) encoding in a visible light communication (VLC) system. In this scheme, the incoming bits are first encoded using the RS encoder, then a set of symbols in the resulting RS codeword are punctured, and the remaining symbols are modulated and mapped onto the OFDM subcarriers. This system is referred to as RS-OFDM-IM. Unlike the traditional subcarrier index modulation (SIM) schemes, the proposed scheme operates based on conveying extra information by inactivating the selected subcarriers, which facilitates simultaneous clipping noise reduction and spectral efficiency enhancement in OFDM-VLC. The bit error rate (BER) and throughput of the proposed technique is theoretically and numerically analyzed. The simulation results show the superiority of the proposed technique as compared to the coded DCO-OFDM without SIM and the classical SIM in OFDM-VLC in a system with practical clipping conditions.**

## I. INTRODUCTION

Demand for high data rates is increasing and the next-generation wireless networks, such as 5G, are predicted to offer high data transmission capacity for new applications. Millions of

[a] Optical Communication Lab, University of Texas at Dallas, Richardson, Texas, USA.

[*] Corresponding author, Email: kamran@utdallas.edu

Internets of Things (IoT) devices and seamless streaming of high-quality videos are predicted to demand a capacity per unit area of 100Mbps for the future indoor spaces [1]. VLC systems can be developed using the available illumination infrastructure inside buildings and can provide a robust data link for short-range applications. VLC can be operated for simultaneous medium illumination and data communication. Further, unlike RF systems, VLC systems do not require a circuitry chain for up/down RF conversion since they work based on direct intensity modulation/detection for data transmission. Orthogonal frequency division multiplexing (OFDM) technique, which is adopted in many recent telecommunication standards, including Long Term Evolution (LTE) and IEEE 802.11x [2], can also be integrated with VLC to overcome the limitation of light-emitting diode (LED) modulation bandwidth and to mitigate the channel dispersion effect in order to increase VLC sum-rate. VLC systems integrated with OFDM and using III-nitride laser diodes have been shown to achieve multi-Gigabit data rates [3, 4]. In OFDM-VLC, the output of the multiplexer needs to be real and positive since the intensity of LED is modulated. Different solutions have been considered to deal with the challenges that are facing OFDM-based VLC, which can be generally categorized into two classes; (1) a DC biased Optical-OFDM (DCO-OFDM) [5], and (2) asymmetrically clipped optical-OFDM (ACO-OFDM) [6]. In ACO-OFDM, information bits are only mapped onto the odd subcarriers and the time-domain signal is clipped at zero to generate a unipolar signal, while in DCO-OFDM, all subcarriers are employed to carry data and then time-domain signal is clipped and biased to generate a non-negative signal. As an alternative to these two dominating schemes, the Unipolar OFDM (U-OFDM) scheme has been proposed in [7], in which the negative and positive samples from the real bipolar OFDM symbol are extracted and transmitted separately by two successive OFDM symbols.

Recently, the concept of subcarrier index modulation (SIM) in OFDM has been proposed,

where indices of subcarriers in OFDM are employed to carry extra information bits. In [8-13], it is shown that SIM changes the spectral efficiency, peak-to-average power ratio (PAPR), bit error rate (BER), power efficiency, and throughput of OFDM system. In this technique, instead of modulating the entire OFDM frame, only a set of subcarriers are chosen to be modulated by M-ary symbols. In addition to these symbols, the combination of the subcarrier indices of the activated subcarriers also carries implicit information. In [8], one index bit is allocated to each subcarrier, and then those subcarriers associated with the subset of the majority bit-value are activated to be mapped by M-ary symbols. In [9], the frequency diversity order and the performance of OFDM with Index Modulation (OFDM-IM) in frequency selective fading channels are investigated. In [10]-[11], the throughput of OFDM-IM is improved by sending symbols of a different constellation on the selected remaining subcarriers, while in dual-mode (DM) OFDM-IM [10], selected and remaining subcarriers are modulated using two different M-ary constellations. In multiple-mode (MM) OFDM-IM multiple distinguishable modes and their permutations are used for symbol modulation [11]. A full review of the different SIM techniques in OFDM can be found in [14].

SIM has been recently extended to VLC to convey extra information bits in a more energy-efficient way. In [15], the system structure of an OFDM-based VLC integrated with SIM is proposed and its BER performance is presented. A hybrid-dimming scheme for both adjusting the illumination level and optimizing the indoor optical channel capacity in SIM-aided VLC is proposed in [16].

Although the energy per active subcarriers is improved by existing SIM in an OFDM with fixed transmission power, the clipping distortion caused by increasing the number of active subcarriers can raise the clipping events and subsequently limit the maximum throughput achievable by these techniques in OFDM-VLC. Hence, the original SIM developed for RF OFDM are not suitable for OFDM-VLC with clipping constraint. Also, majority of the original SIM were originally developed for the uncoded systems and do not fully benefit from the flexibility that can be provided by a channel code. The application of RS codes for improving the reliability of both fiber and VLC is shown and analyzed via numerical and experimental analyses in [17]-[19]. Considering the new dimension introduced by the block code, a new SIM technique that can benefit from this feature must be considered.

Unlike the traditional SIM, which operates based on the activation of selected subcarriers, in RS-OFDM-IM, the SIM attempts to convey implicit information on the indices of a sets of available subcarriers. The process is equivalent to modulating a punctured codeword onto the OFDM frame where the puncturing is conducted by the index modulation in the proposed technique. Since the information is embedded on the nulled subcarriers, the proposed SIM allows a simultaneous data rate increase and clipping noise mitigation in OFDM-VLC.

In [21], the preliminary BER performance of a novel scheme was compared to the uncoded DCO-OFDM with the same bit rate. It was shown that this approach can substantially improve the error floor in OFDM-based VLC systems with clipping constraint. In this work, we extend our preliminary idea and theoretically analyze and explore the potential of the proposed technique in a VLC system with illumination requirements. The contribution of this paper is two-fold.

- A new subcarrier index modulation scheme based on RS encoding is proposed and compared with the traditional subcarrier index modulation employed in a coded OFDM-based VLC and with the DCO-OFDM with RS channel coding. An analytical expression for the lower bound of BER at the output of RS decoder as a function of the effective electrical signal-to-noise and distortion ratio (SNDR) in optical OFDM is derived, and the achievable throughput of the new scheme in OFDM-based VLC under practical clipping conditions and in the presence of Gaussian noise is presented.

- A strategy based on the new scheme is proposed to adjust the dimming level of the OFDM-based VLC. Using the flexibility of RS-OFDM-IM, one can achieve the desire illumination level in the intensity domain while the target SE is maintained. This strategy determines the optimal parameters of RS block code in the transmitter such that the dimming level can be adjusted according to the varying illumination requirement without reducing the bit rate of the system.

The main rationale behind pursuing the proposed approach is to benefit from the new dimension created by the linear block code. Namely, redundancy in the linear block code is used simultaneously for both nonlinear noise mitigation and spectral efficiency. Integration of SIM and RS error-correcting code allows for an effective way of noise mitigation and provides a means to compensate for the bit rate loss caused by the block code.

## II. RS Encoded OFDM-VLC System Model

The technique introduced in this work can be integrated into DCO-OFDM as well as ACO-OFDM VLC systems. However, without the loss of generality, we adopt DCO-OFDM to model the RS-OFDM-IM in this section. The block diagram of RS-OFDM-IM is depicted in Fig. 1. In a DCO-OFDM transmitter with $N_T$ $(even)$ subcarriers, $N_f = \frac{N_T}{2}$ subcarriers are available for individual symbol mapping due to the Hermitian symmetry (HS) requirement. To reduce the computation complexity that can be caused by SIM, $N_f$ subcarriers are divided into G sub-blocks, where each sub-block is allocated $N = \frac{N_f}{G}$ subcarriers ($G$ is selected such that $N$ remains an integer). The information bits are first distributed between sub-blocks and then converted into a non-binary codeword using the Reed Solomon code block, where the length of codewords is determined by the maximum number of subcarriers allocated in each sub-block. The incoming $p_t$ bits in each sub-block are split into two sets; $p_1$ bits are sent to the index selector to be conveyed by SIM while $p_2$ bits are transmitted via $M$-ary constellation points.

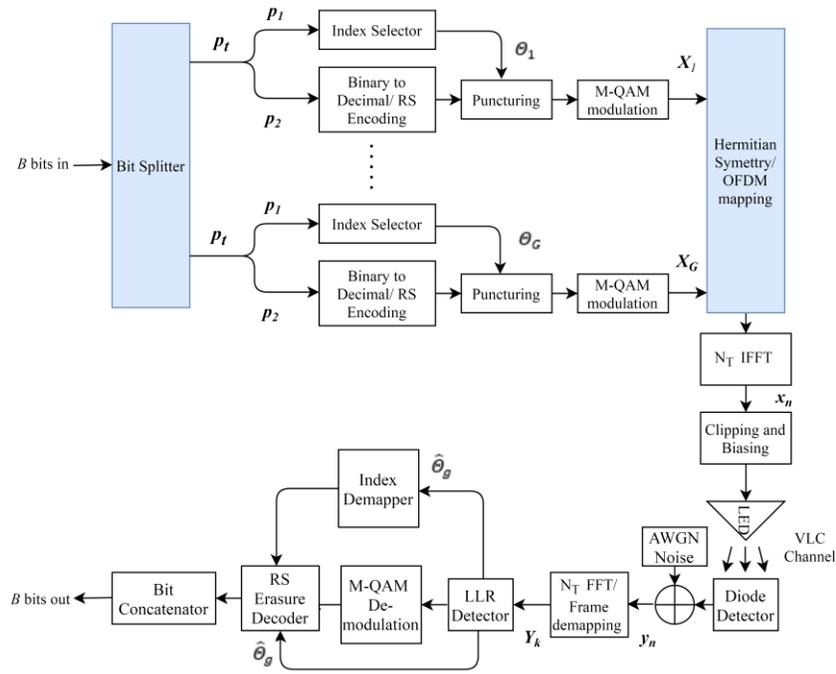

Fig. 1. Block diagram of RS-OFDM-IM.

Considering an RS code over the Galois field $GF(2^m)$ with code length $N = 2^m - 1$, the non-binary message vector of size K which carries $p_2 = Km$ information bits is encoded to generate a codeword with $N$ RS symbols. The RS encoder can be represented by $RS\ (N, K, d = N - K + 1)$, with the code rate $r_c = \frac{K}{N}$. The encoder uses the message vector in its generator polynomial to generate codewords with a minimum distance of $d = N - K + 1$. As an erasure code, the RS decoder can detect and correct combinations of $\xi$ errors and $S$ erasures, if $2\xi + S \leq d - 1$. Given the linear block code $\mathcal{B} \in \mathcal{M}^N$, a set with $|\mathcal{B}|$ number of $N$-tuples over the alphabet $\mathcal{M} = \{0,1,\ldots,M-1\}$ with hamming distance $d$ can generate $M^K$ distinct codewords which will be used in the scheme shown in Fig. 1 for the modulation of OFDM sub-blocks. Let $\phi = \{\omega_g | g =$

$1, \dots G; \omega_g \in \mathcal{B}\}$ be the set of codewords used for the OFDM frame mapping, where $\omega_g$ is the codeword allocated to the sub-block $g$. The index selector uses $p_1 = \left\lfloor log_2 \binom{N}{S} \right\rfloor$ bits to select the puncturing vector $\Theta_g = a_i$ from a predefined set $\mathcal{A} = \{a_1, a_1 \dots, a_{2^{p_1}}\}$, where the elements of $a_i$ determine the symbol indices in the codeword that will be punctured, and their corresponding subcarriers will be set to inactive. Considering the puncturing vector of the $g^{th}$ sub-block given by

$$\Theta_g = \{\theta_{g,1}, \dots, \theta_{g,S}\} \tag{1}$$

where $\theta_{g,s} \in \{1, \dots, N\}$, and after nulling subcarriers in $\Theta_g$, the remaining subcarriers in the sub-block will be modulated using the intact part of the codeword. Now, the symbol vector $\Omega_g \in \mathbb{C}^{N \times 1}$ can be expressed as

$$X_g = [X_1^g, \dots, X_N^g] \tag{2}$$

where $X_i^g = 0$ when $i \in \Theta_g$, and for $i \notin \Theta_g$, $X_i^g = \mathcal{U}_m \in \{M - QAM\}$. After the formation of symbol vectors for every sub-block and applying HS, the frequency-domain OFDM block is created by concatenating the symbol vectors of all G sub-blocks. Upon applying the HS, the OFDM frame with total $N_T$ subcarriers in the frequency-domain is represented by

$$X = [0, X_1^1, \dots, X_N^1, \dots, X_1^G, \dots, X_N^G, 0, X_N^{G^*}, \dots, X_1^{G^*}, \dots, X_N^{1^*}, \dots, X_1^{1^*}] \tag{3}$$

where $X_i^{j^*}$ indicates the complex conjugate of symbol $X_i^j$. The extended frequency-domain block is then converted to the time-domain OFDM symbols by $N_T$-points Inverse Fast Fourier Transform (IFFT) operator given by (for ease of notation, $X_k$ is used to denote the $k^{th}$ symbol in the frequency- domain block)

$$x_n = \frac{1}{\sqrt{N_T}} \sum_{k=1}^{N_T} X_k e^{\frac{2\pi j n k}{N_T}} \tag{4}$$

For large IFFT points, i.e., $N_T > 64$, the output is known to be Gaussian distributed with zero mean and standard deviation $\sigma_x$, i.e., $x_n \sim N(0, \sigma_x^2)$.

To make $x_n$, $1 \leq n \leq N_T$ transmittable in an LED with by $\Lambda = [I_L, I_H]$ dynamic range, bipolar time-domain signal is converted to a unipolar signal. This can be represented by

$$x_{n,t} = x_{n,c} + I_{bias} ; \qquad (5)$$

$$x_{n,c} = clip(x_n) = \begin{cases} u_{lower} & x_n \leq u_{lower} \\ x_n & u_{lower} < x_n < u_{upper} \\ u_{upper} & x_n \geq u_{upper} \end{cases} \qquad (6)$$

with $u_{lower} = (I_L - I_{bias})$ and $u_{upper} = (I_H - I_{bias})$, denoting the lower bound and upper bound, respectively, which are set relative to the standard deviation of $x_n$ by constant coefficients $\lambda_1$ and $\lambda_2$ [22], i.e. $u_{upper} = \lambda_1 \sigma_x$, $u_{lower} = \lambda_2 \sigma_x$. The probability of clipping at both ends of LED's are calculated by $P_l = 1 - Q(\lambda_2)$ and $P_u = Q(\lambda_1)$, respectively, where $Q(a) = \frac{1}{\sqrt{2\pi}} \int_a^\infty \exp\left(-\frac{b^2}{2}\right) db$. The clipped signal is usually modeled using Bussgang theorem as [22]

$$x_{n,c} = \beta x_n + z_n \qquad (7)$$

where $\beta$ is the attenuation factor given by $\beta = E\{x_{n,c} x_n\}/E\{x_n^2\}$ and $z_n$ is the clipping noise, which is assumed to be uncorrelated with the signal $x_n$. Using (5)- (7), the statistics of truncated Gaussian signal and clipping noise can be calculated by

$$\sigma_{x_c}^2 = \sigma_x^2 [Q(\lambda_2) - Q(\lambda_1) + \phi(\lambda_2)\lambda_2 - \phi(\lambda_1)\lambda_1 + (1 - Q(\lambda_2))\lambda_2^2 + Q(\lambda_1)\lambda_1^2] - E[x_{n,c}]^2 ,$$

$$E[x_{n,c}] = \sigma_x [\phi(\lambda_2) - \phi(\lambda_1) + (1 - Q(\lambda_2))\lambda_2 + \lambda_1 Q(\lambda_1)] ,$$

$$\sigma_z^2 = \sigma_{x_c}^2 - \beta^2 \sigma_x^2 \qquad (8)$$

where $\phi(a) = \frac{1}{\sqrt{2\pi}} exp\left(-\frac{a^2}{2}\right)$, and $\sigma_{x_c}^2$ is the variance of the truncated Gaussian signal. After adding the cyclic prefix to the OFDM symbols, the signal is transmitted over the linear time-invariant baseband VLC channel. The received signal using direct-detection is give

$$y_n = (\beta x_n + z_n + I_{bias}) * h_n + w_n \qquad (9)$$

where $h_n$ denotes channel impulse response, $w_n \sim \mathcal{N}(0, \sigma_w^2)$ is a Gaussian noise, and * denotes the convolution operation. The received frequency-domain symbols are given by

$$Y_k = \beta H_k X_k + H_k Z_k + W_k, \quad 1 \leq k \leq \frac{N_T}{2} \tag{10}$$

where $Z_k$ and $H_k$ are clipping noise and the channel overall gain at $k$-th subcarrier, respectively. The receiver splits the frequency-domain symbols into G sub-blocks, and detect the indices of nulled subcarriers and estimate the puncturing vectors to decode $p_1$ information bits. Also, by demodulating the symbols of the active subcarrier in each sub-block, its codeword is recovered and sent to the RS decoder for decoding of $p_2$ information bits that are transmitted by active subcarriers. Two detection algorithms, *i.e.,* Maximum Likelihood (ML) and Log-Likelihood Ratio (LLR), can be used for null indices detection. However, due to the fairly large number of subcarriers and the M-ary modulation, the LLR detection with lower complexity is a more suitable candidate. Therefore, the LLR is employed in Fig. 1. The LLR detector calculates the ratio of a posteriori probabilities of the complex symbols in the frequency-domain based on the fact that the inactive subcarrier is null, while active subcarriers are modulated by M-QAM constellation points. For subcarrier $k$ in the $g^{th}$ sub-block, this ratio is given by

$$\alpha_g(k) = \ln \frac{\sum_{m=1}^{M} P(X_k^g = \mathcal{U}_m | Y_k^g)}{P(X_k^g = 0 | Y_k^g)} \tag{11}$$

where $Y_k^g = Y_{N(g-1)+k}$ is the received signal on the $k^{th}$ subcarrier of the $g^{th}$ sub-block. Using Bayesian rule, (13) can be simplified to

$$\alpha_g(k) = \ln S - \ln(N - S) + \frac{|Y_k^g|^2}{\sigma_W^2} + \ln(\sum_{i=1}^{M} \exp(-\frac{1}{\sigma_w^2}|Y_k^g - H_k^g \mathcal{U}_m|^2)) \tag{12}$$

The simplification in (12) is based on the following relations: $P(X_k^g = 0) = \frac{S}{N}$ and $\sum_{m=1}^{M} P(X_k^g = \mathcal{U}_m) = \frac{N-S}{N}$. The LLR detector searches through all subcarriers and for each sub-block it finds indices of $S$ subcarriers with the lowest LLR values. Given the sorted LLR values as $\alpha_g(q_1) > \alpha_g(q_2) > \cdots > \alpha_g(q_N)$, the puncturing vector is then estimate as

$$\hat{\Theta}_g = \{q_{N-S+1}, q_{N-S+2}, \ldots, q_N\} \quad \Theta_G \quad (13)$$

The index de-mapper uses the transmitter-receiver common look-up table to decode $b_1$ bits used in the transmitter for index selection and informs the decoder about the indices of the codeword erasure. The decoder performs the error-correction decoding using the remaining code redundancy. Finally, the recovered bits from index de-mapper and RS decoder collected from all $G$ sub-blocks are concatenated to form the output information vector in the receiver

## III. SIM IN RS ENCODED OFDM-VLC

In this section, first, a strategy based on RS-OFDM-IM is proposed to improve BER when the dimming level of LED changes for illumination purposes.

### A. Dimming level control strategy

Considering the RS encoded modulation, the size of symbols can be reduced such that the resulted signal in the time-domain experiences fewer clipping events. However, the coding rate and the number of punctured symbols in codewords should be determined such that the system's target SE can be achieved.

A strategy based on the flexibility of coding parameters can be developed to reduce the clipping noise when the VLC dimming level alters. This strategy aims at controlling the probability of clipping by choosing the right values for the message length and number of punctured symbol that can achieve a target SE and also lead to a smaller magnitude in the envelope of the time-domain signal. Considering an LED with dynamic range $\Lambda = [I_L, I_H]$, the dimming level is calculated as $\eta = \frac{I_m - I_L}{I_H - I_L}$ ($0 \leq \eta \leq 1$), where $I_m$ is the average intensity of the LED. Rewriting the upper and lower clipping bounds as $I_H - I_{bias} = \lambda_1 \sigma_x$ and $I_L - I_{bias} = \lambda_2 \sigma_x$, respectively, $I_m$ is given by

$$I_m = E\{x_{k,t}\} = \sigma_x[\lambda_1 Q(\lambda_1) + \lambda_2 Q(-\lambda_2) + \phi(\lambda_2) - \phi(\lambda_1)] + I_{bias} \tag{14}$$

where $E\{\}$ is the ensemble average of the enclosed, and $\phi(u) = \frac{1}{\sqrt{2\pi}} e^{-\frac{u^2}{2}}$. Assuming the power normalized symbols on active OFDM subcarriers, the standard deviation of the time-domain signal, after inactivating $S$ subcarriers in each sub-blocks according to (4) is now given by

$$\sigma_x'' = \sqrt{\frac{N_T - 2GS - 2}{N_T}} \sigma_x \tag{15}$$

It can be seen that by nulling a set of subcarriers using SIM, the time-domain envelope of IFFT multiplexer is controlled. In a VLC system with fixed modulation bandwidth and $N_T$ total subcarriers and $G$ independent sub-blocks, the SE of RS-OFDM-IM with message length $K$ and assuming $M$-QAM employment is

$$SE_{RSO}(M, K, S) = \frac{G}{N_T}(K \log_2 M + \lfloor \log_2 C(N,S) \rfloor) \text{ (bit/s/Hz)} \tag{16}$$

where $C(a,b) = \frac{a!}{b!(a-b)!}$ denotes the binomial coefficient. It should be noted that the size of the RS code alphabet should not be greater than the number of points in the M-ary constellation used for symbol modulation, *i.e.*, $N \leq M$.

In a VLC system with fixed bandwidth and subcarriers, RS-OFDM-IM can achieve a SE similar to DCO-OFDM with shorter message length and smaller code rate. Assume a DCO-OFDM modulation, with $RS\ (N, K', d = N - K' + 1)$ encoder, the SE of DCO-OFDM is given by $SE_{DCO}(M, K') = G \frac{K'}{N_T} \log_2^M$. The set of parameters that can achieve the equal or greater SE in RS-OFDM-IM is represented by

$$\{(K_p, S_p)\} = \{(K, S) \mid SE_{RSO}(M, K, S) \geq SE_{DCO}(M, K')\} \tag{17}$$

Given that any 2-tuple in the set $\{(K_p, S_p)\}$ will satisfy the required data rate, the larger value of $S_p$ will lead to greater $\lambda_1$ and $\lambda_2$ values. Stretching of clipping boundaries by increasing the number of punctured symbols in each sub-block's codeword will reduce the probability of clipping and will keep the large signal in the intensity domain intact. This impact will become more pronounced in a VLC system with a highly varying illumination level. For quantitative analysis, we consider an OFDM-based VLC with $N_t=1024$ subcarriers and a LED with the dynamic range given by $\Lambda = [0,5]$. Information in this system is encoded over $GF(2^5)$ to generate codewords of length 31, and then are modulated by 32-QAM and then mapped over subcarriers in $G = 16$ sub-blocks. Table 1 shows the code parameters for two schemes and their corresponding clipping bounds at different dimming levels.

Table.1 Clipping bounds in RS-OFDM-IM and DCO-OFDM in VLC with $\Lambda = [0,5]$ physical constraint.

|  | $(K,S)$ | upper-lower bounds | $\eta$: 0.1 | $\eta$: 0.3 | $\eta$: 0.5 | $\eta$: 0.7 | $\eta$: 0.9 |
|---|---|---|---|---|---|---|---|
| RS-OFDM-IM | (20,3) | $\lambda_2$ | -0.95 | -1.75 | -2.67 | -3.58 | -4.39 |
|  |  | $\lambda_1$ | 4.39 | 3.58 | 2.67 | 1.75 | 0.95 |
|  | (19,5) | $\lambda_2$ | -0.96 | -1.80 | -2.77 | -3.74 | -4.6 |
|  |  | $\lambda_1$ | 4.6 | 3.74 | 2.77 | 1.80 | 0.96 |
|  | (18,7) | $\lambda_2$ | -0.99 | -1.86 | -2.88 | -3.91 | -4.79 |
|  |  | $\lambda_1$ | 4.79 | 3.91 | 2.88 | 1.86 | 0.99 |
| DCO-OFDM | (22) | $\lambda_2$ | -0.93 | -1.6 | -2.54 | -3.3 | -4.15 |
|  |  | $\lambda_1$ | 4.15 | 3.3 | 2.54 | 1.6 | 0.93 |

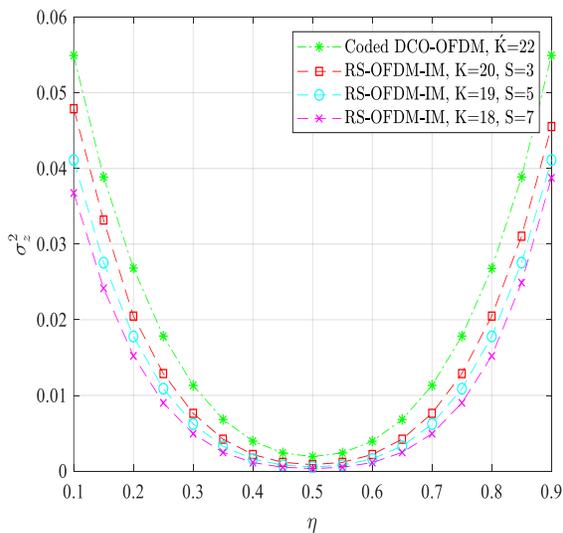
Fig. 2. Clipping noise power in RS-OFDM-IM vs. RS Encoded DCO-OFDM.

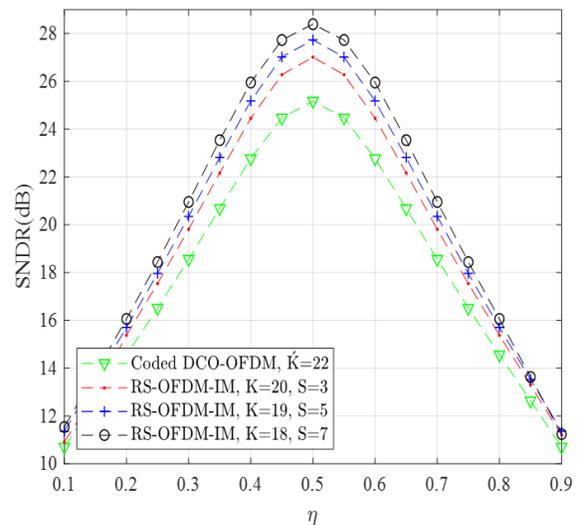
Fig. 3. average SNDR comparison in dimmable OFDM-VLC at SNR = 30 dB.

A SE close to the target SE can be achieved by all different block code parameters that are shown in the table.1; however, as the number of null subcarriers increases in sub-blocks, the time-domain OFDM signal will experience a less severe clipping noise. Fig. 2 show the reduction in clipping noise power for different puncturing numbers. Fig.3 shows the level of change in the average Signal to Noise and Distortion Ratio (SNDR) of time-domain signal, calculated by $SNDR = \frac{\beta^2 \sigma_x^2}{\sigma_z^2 + \sigma_{AWGN}^2}$ when the received signal has a SNR of 30 $dB$. Fig.2 and Fig.3 prove the improvement in the conditioning of the transmitted signal in RS-OFDM-IM compared to DCO-OFDM with a similar SE.

Considering an indoor medium with both line-of-sight and non-line-of-sight transmission path, the impact of channel in (10) cam be compensated by the receiver equalization, e.g. Zero Forcing (ZF) before making decisions on the subcarrier's status and their symbols. Hence, $H_k$ in (9) will become a factor in the equalization process and will result in equal SNR penalty in optical OFDM. Here we assume gain on different subcarriers to be 1 for simplicity. Note that no fading is assumed here as fading is not an issue in medium related to VLC for which the RS-OFDM-IM is primarily considered and developed. The SNDR of frequency-domain on *k*th active subcarrier when ZF equalization is used can be calculated by

$$SNDR_k = \frac{\beta^2 \sigma_x^2}{\sigma_z^2 + \left(\frac{A_{DC}\sigma_{AWGN}^2}{H_K^2}\right)} = \frac{\beta^2}{\left(\frac{\sigma_z^2}{\sigma_x^2}\right) + \left(\frac{A_{DC}}{SNR_l H_k^2}\right)} \tag{18}$$

where $SNR_l = E\{x_n^2\}/\sigma_{AWGN}^2$ is the electrical SNR of the original unclipped signal $x_n$. When the clipping is mild, the average power of the clipped signal is approximately the same as the original signal, $SNR_l$ , and it can be rewritten as $SNR_l \approx SNR_t - I_{bias}^2/\sigma_{AWGN}^2$, where $SNR_t$ denotes the electrical SNR of LED's transmitting signal. And $A_{DC} = (\sigma_x^2 + I_{bias}^2)/\sigma_x^2$ is the attenuation of electrical power due to the DC biasing [22]. An erroneous decision on indices in (12) will lead to error in the SIM embedded bits and also erasures of wrong symbols in the received codeword, which will further reduce the RS block code error-correction capability. The number of erroneous erasures in a codeword of the $g^{th}$ sub-block is given by

$$\pi_g = \frac{\|X_g - \hat{X}_g\|_0}{2} \tag{19}$$

where $\|A\|_0$ is the $l_0$ norm of $A$ and $\hat{X}_g$ is an estimate of $X_g$. The average number of symbol errors and erroneous erasures in the codeword is $\Delta_g = \pi_g + NP_{SER}$. The symbol error probability, $P_{SER}$, of data transmitted using a square M-QAM modulation in OFDM with an average subcarrier SNDR given by (18) is given by

$$P_S = 2\left(1 - \frac{1}{\sqrt{M}}\right) erfc\left(\sqrt{\frac{3}{2(M-1)} SNDR_k}\right) - \left(1 - \frac{2}{\sqrt{M}} + \frac{1}{M}\right) erfc^2\left(\sqrt{\frac{3}{2(M-1)} SNDR_k}\right) \tag{20}$$

Note that part of the redundancy that was punctured according to SIM cannot be used for codeword recovery. The remaining redundancy which is to be used by the decoder has to be sufficient for correcting both false erasure and erroneous symbol recovery. Considering that the active subcarriers in each sub-block in RS- OFDM-IM scheme belong to a common codeword, calculating the probability of correction when the LLR detector is used can be rather difficult due to the dependence between separate subcarriers. Considering a system with a large block code size, the frequency-domain symbols can be considered approximately uncorrelated, the probability of correct detection of active subcarriers in each sub-block can be approximated. Let the set of indices of active subcarriers be $\Psi_g = \{\psi_{g,1}, \dots, \psi_{g,N-S}\}$. Also, let $\Psi_g$ be a set that is a complement to $\Theta_g$, i.e., $\Psi_g \cup \Theta_g = \{1,2,\dots,N\}$. Now, a correct estimation of the puncture vector requires the receiver to distinguish between every active subcarrier in the set $\Psi_g$ and every inactive subcarrier in $\Theta_g$. Without the loss of generality, we first consider the probability of correctly distinguishing between the first active subcarrier $\psi_{g,1}$ and the first inactive subcarrier $\theta_{g,1}$ in the $g^{th}$ sub-block whose symbols are given by $Y_g^{\psi_{g,1}} = \beta X_g^{\psi_{g,1}} + Z_g^{\psi_{g,1}} + W_g^{\psi_{g,1}}$ and $Y_g^{\theta_{g,1}} = Z_g^{\theta_{g,1}} + W_g^{\theta_{g,1}}$, respectively. Using the results from [23] and [24], the probability of correctly distinguishing between $\psi_{g,1}$ and $\theta_{g,1}$ is given by

$$P_c = 1 - \frac{1}{2M}\sum_{m=1}^{M} e^{-\frac{u_m^2}{2(\sigma_W^2+\sigma_Z^2)}} \tag{21}$$

with $\mathcal{U}_m$ denoting the constellation points in the M-QAM constellation. Now, the probability of correctly distinguishing all active subcarriers from all inactive subcarriers can be calculated by

$$P_{cd} = (1 - \frac{1}{2M}\sum_{t=1}^{M} e^{-\frac{u_m^2}{2(\sigma_W^2+\sigma_Z^2)}})^{s(N-s)} \tag{22}$$

When the active subcarrier in a sub-block are detected correctly we will have $r = \|\Theta_g - \hat{\Theta}_g\|_0 = 0$. In this case, subcarrier index demolation will have zero error, and all the erroneous bits are only attributed to demodulation of symbols on active subcarriers. The symbol error probability at the decoder input is equal to that of M-QAM demodulation, i.e., $P_S^{in}(r=0) = P_{SER}$. Now, the symbol error probablity at the decoder's output is calculated by

$$P_{S,c}^{out} = \sum_{v=t+1}^{N} \frac{v}{N}\binom{N}{v} P_S^v (1-P_S)^{N-v} \tag{23}$$

Where $t = \lfloor\frac{N-K-s}{2}\rfloor$ denotes the number of correctable errors. In the case of incorrect indix detection, the puncturing vector $\bar{i}$ may be decoded incorrectly as one of the remaining $2^{p_1} - 1$ vectors in $\mathcal{A}$ with equal probability of $\frac{1-P_{cd}}{2^{p_1}-1}$. But the number of wrong errasures in a codeword will depends on how many different element the actual and estimated puncturing vectors have. Therefore, all possible number of differences in $\Theta_g = a_i$ and $\hat{\Theta}_g = a_{\hat{i}}$ should be considered since decoder's particular performance depends on the exact number of difference between these two vectors. Considering $r = \|\Theta_g - \hat{\Theta}_g\|_0 \neq 0$, symbol error probability at the decoder input is given by

$$P_S^{in}(r) = \frac{r+(N-S-r)\times P_S^{in}(0)}{N-S} \qquad (24)$$

By calculating the different realizations that will lead to $r$ number of wrong erasures in the codeword, the symbols error proability at decoder's output can be approximated by

$$P_{S,inc}^{out} \approx \frac{1}{2^{p_1}-1}\sum_{r=1}^{S}\sum_{v=\theta}^{N-r}\binom{S}{r}\binom{N-S}{r}\binom{N-r}{v}\frac{v+r}{N}(P_S^{in})^v(1-P_S^{in})^{N-r-v} \qquad (25)$$

Where $\theta = \max(t-r+1, 0)$. Using (23)-(27), the RS decoder's symbol error probability is calculated as

$$P_{S,out} = P_{cd}P_{S,c}^{out} + (1-P_{cd})P_{S,inc}^{out} \qquad (26)$$

Now, The upper and lower bounds of decoder's bit error probability, $P_b$ in both cases can be calculated from its symbol error probability i.e. $\frac{P_{S,out}}{m} \leq P_b \leq P_{S,out}$.

To calculate the total error rate, the probability of error for the information carried by the SIM should also be considered. The information conveyed SIM is in error only if the subcarriers' status are detected incorrectly. Hence, subcarrier index demoultion contributes to the total BER when $\|\Theta_g - \hat{\Theta}_g\|_0 \neq 0$. Note that different wrong puncturing vectors could lead to different number of erroneous bits in the subcarrier index demoulation. However, since all the wrong puncturing vectors are equally probable to be selected ($\frac{1}{(2^{p_1}-1)}$), and because there is a total of $\binom{p_1}{t}$ realizations that will have $t$ erroneous bits, the average BER in subcarrier index demodulation can be approximated by

$$P_b^{SIM} \approx \frac{1}{p_1(2^{p_1}-1)}\sum_{t=1}^{p_1} t\binom{p_1}{t} \qquad (27)$$

Recalling that each message contains $p_2 = Km$ information bits, the lower bound of total bit error probability in RS-OFDM-IM is calculated by

$$P_b \geq P_{cd}\frac{P_{S,c}^{out}\times K}{P_1+P_2} + (1-P_{cd})\frac{(P_{S,inc}^{out}\times K + P_1 P_b^{SIM})}{P_1+P_2} \qquad (28)$$

The bound can be calculated by substituting (23), (25), and (27) in (28). It should be noted that the probability of successful detection of non-signal subcarriers and RS decoding of signal subcarriers are not independent of the specific sequence of QAM symbols, , and the expression in (28) is essentially computed by using the product of the average of two error rates, and it gives an approximation of the actual BER in the underlying systems. However, the results in the next section show its closeness for various encoding and modulation configurations. Using the overall BER and SE of VLC, we can also calculate :

$$R_b = (1 - P_b)SE_{RSO}(M,K,S) \qquad (29)$$

where $R_b$ represent the throughput (bps/Hz), which accounts for the successful transmission rate per bandwidth that is achievable by RS-OFDM-IM in a VLC with the clipping constraint.

### B. Complexity

The computational complexity in the receiver of RS-OFDM-IM is compared to the original OFDM system. The computational complexity of LLR detection in (14) for each subcarrier will be in the order of $\mathcal{O}(M)$, which is the same as the detection complexity in the original OFDM. Considering the $2^{p_1}$ possible realization for puncturing vector $\Theta_g$, and $N - s$ active subcarriers in each sub-block and M-QAM employment, the ML detection in RS-OFDM-IM will entail $\mathcal{O}(2^{p_1} M^{N-s})$ complex multiplication per sub-block. In a system with encoder and decoder that operates over a fairly large finite field and also use a high order of QAM modulation, LLR is a better option as it can achieve a trade-off between detection accuracy and complexity [9]. While the classical RS decoding algorithm in RS(N,K,N-K+1) code yields $\mathcal{O}(N^2)$ computational complexity, the recent works on reformulating the basis of syndrome polynomial and using FFT algorithm for decoding procedure have shown that the error-correction decoding algorithm over binary extension field GF($2^m$) can be reduced to $\mathcal{O}(Nlog(N-K) + (N-K)(log_2^{(N-k)})^2)$ [25], while the fastest algorithm for the case of erasure decoding can achieve $O(NlogN)$ complexity [26].

IV. SIMULATION RESULTS

Given the discussion in the previous section, the coding based SIM can be used for obtaining a better SE by sending extra on non-signal subcarriers, and improving the system BER by mitigating the clipping noise. We compare the BER performance of RS-OFDM-IM with RS encoded DCO-OFDM and RS-encoded OFDM with classical index modulation in a Gaussian channel. First, the accuracy of the analytical expression for BER in RS-OFDM-IM with clipping and channel noise is examined. Then, we compare the BER performance of the three technqiues mentioned above in both single-sided and double-sided clipping conditions. Furthermore, the performance of the proposed strategy in improving the throughput of OFDM-VLC subject to dimming level variation is presented and analyzed.

It was discussed that inactivating an increased number of subcarriers in a sub-block can further reduce the clipping noise. However, this increase can compromise RS code's error-correction capability. The overall effect of this increase on the BER is shown as a function of received SNR in Fig. 4. A system with $N_t=1024$ is considered, where from 512 available subcarriers before HS, 498 subcarriers are divided into $G = 16$ sub-blocks, with $N = 31$ subcarriers each and 16 overall subcarriers are left for guard band, and $RS\ (31, K, d = 31 - K + 1)$ encoder and 32-QAM modulation are employed for symbol modulation. The BER performance at two different message lengths $K = 19$ and $25$ with corresponding coding rates $r_c \approx 65\%$ and $80\%$, respectively, are presented. The SE of this OFDM system can be calculated as $\frac{16\left(\left\lfloor log_2\binom{31}{S}\right\rfloor + Klog_2^{32}\right)}{1024}$, and by increasing $S$, the size of information that are embedded on non-signal subcarriers are adjusted. Considering three different values of puncturing $S = 2, 4$ and $6$ when $K = 19$, the SE of OFDM system yields 1.60, 1.70, and 1.78 (bit/s/Hz), respectively, while for $K =$

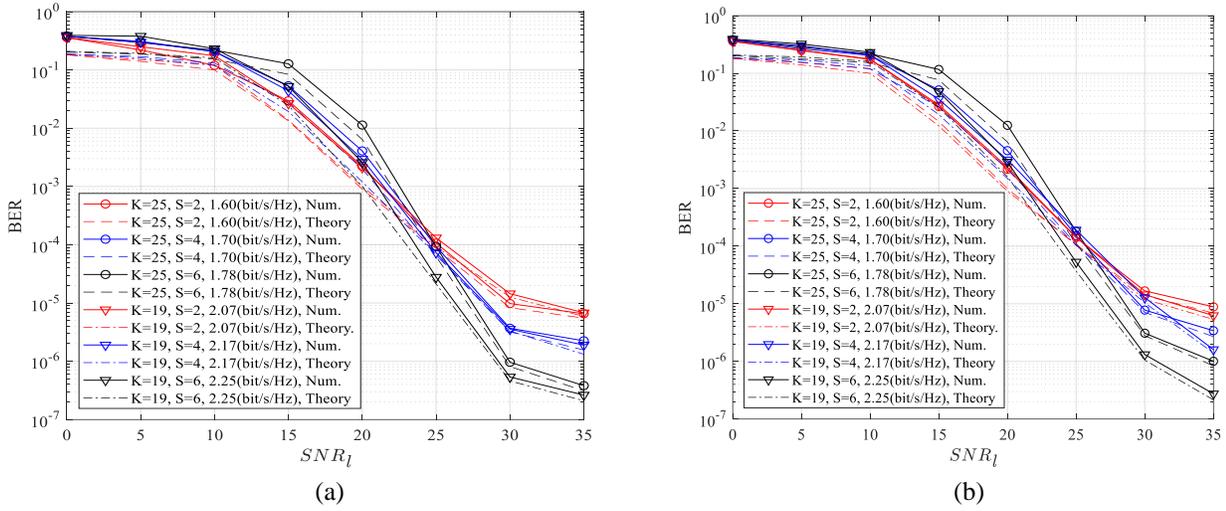

Fig.4. BER of RS-OFDM-IM in AWGN channel with varying puncturing number; (a) Double-sided clipping, (b) Single-sided clipping.

25 it gives 2.07, 2.17, and 2.25 (bit/s/Hz), respectively. Figs. 4 (a) and (b) show the effect of increasing $S$ in double-sided and single-sided clipping conditions. In double-sided clipping, bounds are assumed to be $u_{lower} = -2\sigma_x$, $u_{upper} = 2\sigma_x$ where for single-sided clipping, it's $u_{lower} = -1.8\sigma_x$, $u_{upper} \rightarrow \infty$; and $E\{x_n^2\} = \sigma_x^2$ is the power of unclipped signal when no puncturing is applied and all subcarriers are active.It can be seen that in both clipping conditions, at a particular coding rate a system with greater puncturing number yields a better BER. Note than the error floors in Fig. 5 are due to the clipping noise which becomes dominant over channel noise at high SNR levels. Therefore as the number of punctured symbols increases, the reduction in clipping distortion on OFDM subcarriers outweighs the loss in the RS code's error-correction capability and yields a lower error floor. Also, reducing the coding rate in single-sided clipping condition has more notable effect in improving the overall BER, while in double-sided clipping with more severe clipping noise, the overall BER of the smaller coding rate is only slightly better for a particular puncturing number. This is because the erroneous bits are mainly related to the LLR detection of $p_1$ bits that are carried out by SIM. The correct recovery of these bits is mostly determined by the number of non-signal subcarriers. Hence, the BER to a large extend is independent of the coding rate under severe clipping. It should be noted that the improvements in SE and BER come at the expense of more computations in the system since the size of the lookup table and the number of searches during LLR detection are proportional to $S$. It can be seen that the theoretical bound given in (28) shows a good match to the numerical values especially at higher SNR.

To see the advantage of the proposed SIM in RS-OFDM-IM, its performance is also compared to the RS coded OFDM with the traditional index modulation (OFDM-IM), where only active subcarriers are used to transmit both index embedded infromation and QAM symbols.

For the sake of a fair comparison, we assume that the codeword size and coding rate are the same in both RS-OFDM-IM and RS coded OFDM-IM. The number of employed subcarriers is determined such that a common SE can be achieved by both schemes. Considering a system with 1090 available subcarriers, and 545 independent subcarriers before applying HS, a total of $G = 16$ sub-blocks can be formed with 34 subcarriers in each. The codewords are generated over Galois field $GF(2^5)$ with a length of 31, modulated by 32-QAM, and then mapped on the OFDM frame and transmitted over a Gaussian channel. Suppose an $RS(31, K, 31-K+1)$ encoder in commonly employed in both schemes. In traditional OFDM-IM, the 31 symbols of each codeword are mapped on their associated subcarriers according to SIM in each sub-block. After modulating all 16 sub-blocks, the time-domain is generated and then clipped according to the optical front-end clipping bounds. At the receiver, given that the symbols of the codeword order during index modulation is preserved, the LLR detector can reconstruct the valid codewords without introducing a displacement error. The number of information bits that can be transmitted in one sub-block in traditional OFDM-IM is $p'_t = \left\lfloor log_2 \binom{34}{31} \right\rfloor + K log_2^{32}$. In RS-OFDM-IM with the same coding rate, VLC can transmit $p_t = \left\lfloor log_2 \binom{31}{S} \right\rfloor + K log_2^{32}$ bits per sub-block. Assuming $S = 3$, the same number of information bits and SE can be achieved by both schemes. The BER performances of the two schemes as a function of $SNR_l$ in a VLC with double-sided ($u_{lower} = -2\sigma_x$, $u_{upper} = 2\sigma_x$) and single-sided ($u_{lower} = -1.8\sigma_x$, $u_{upper} \to \infty$) clipping are shown in Fig. 5(a) and Fig. 5(b), respectively. At two coding rates, $r_c = 54\%$ (K=17) and $r_c = 74\%$ (K=23), the total BER of the systems are presented.

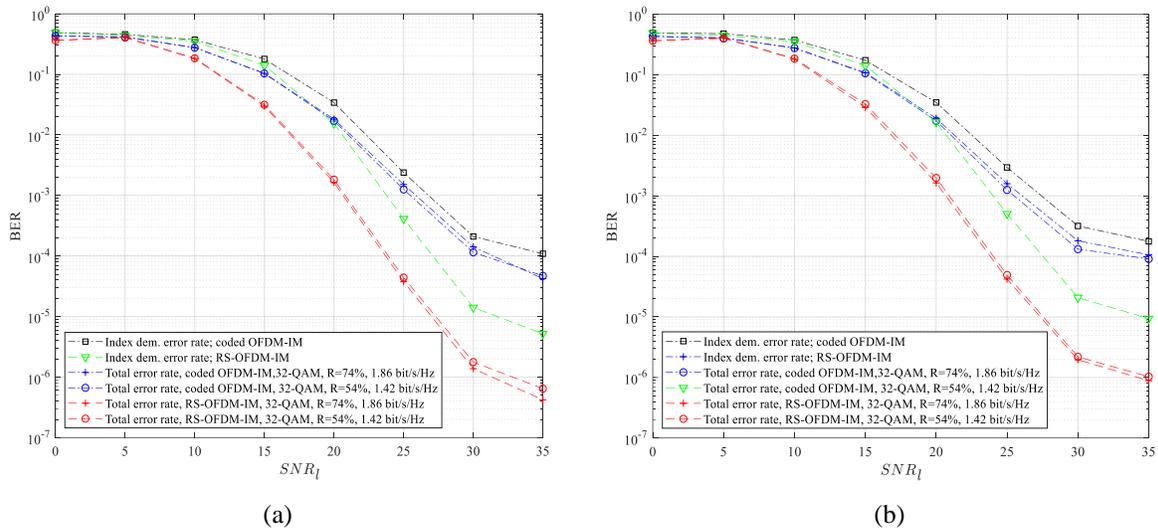

Fig.5 RS coded traditional OFDM-IM vs. RS-OFDM-IM in AWGN with codewords over GF ($2^5$) and 32-QAM employment ; (a) Double-sided clipping, (b) Single-sided clipping.

Although the BER of both schemes plateau in both clipping conditions, the error floor in RS-OFDM-IM is significantly smaller. Note that the amount of information transmitted using SIM and the total bitrate are the same in both schemes. But the index demodulation error, which accounts for errors in bits transmitted by the SIM is smaller in RS-OFDM-IM. Since information is embedded in indices of a smaller number of subcarriers in RS-OFDM-IM as compared to the traditional OFDM-IM (31< 34), the likelihood of an erroneous index detection due to clipping or channel noise is less in this scheme. Also, due to the increased number of non-signal subcarriers and reduced clipping noise, the LLR detector has better accuracy in indices estimation which improves the total BER performance in RS-OFDM-IM. The superiority of RS-OFDM-IM in achieving the same bitrate while substantially improving BER shows its advantage in comparison to the traditional SIM scheme in a coded OFDM system.

We also compare RS-OFDM-IM against the basic OFDM-based VLC in the absence of SIM. DCO-OFDM, which is well known for its superior SE when compared to ACO-OFDM, is used for the comparison. In RS encoded DCO-OFDM, all available subcarrier are modulated and data are conveyed by QAM symbols. The BER performances of the two systems are compared and presented in Fig. 6. In a system with $N_t$=1024 total subcarriers, and 512 available subcarriers before applying HS, 496 subcarriers are considered for data transmission. Information bits in DCO-OFDM are encoded in the frequency-domain by a non-binary encoder $RS(31, K', 31-K'+1)$ over $GF(2^5)$. Symbols of 16 codewords are then modulated by 32-QAM constellation and then mapped onto the data subcarriers. The SE of DCO-OFDM is given by $SE_{DCO}(K') = (16\,K'log_2^{32})/1024$. In RS-OFDM-IM, the same 496 available subcarriers are divided and used for forming 16 sub-blocks with 31 subcarriers, where each sub-block is modulated by codewords generated using an $RS(31, K, 31-K+1)$ encoder. Considering $S$ puncturing in each codeword, SE is given by $SE_{RSO}(M,K,S) = 16(Klog_2^{32} + \lfloor log_2\,C(31,S)\rfloor)/1024$. The modulation parameters are adjusted such that a similar SE can achieved by RS-OFDM-IM and DCO-OFDM. Since in RS-OFDM-IM information is conveyed by both active and inactive subcarriers in, the length of the message at the input of the encoder could be smaller ($K' > K$) and it would still achieve a similar bitrate with a lower coding rate. Using (19), the $\{K_p, S_p\}$ sets that generate an SE equal or greater than that of coded DCO-OFDM at two message length of $K' = 26$ and $K' = 22$ are calculated. For the first case, we have $SE_{DCO}(32,26) = 2.03$ (bit/s/Hz) in coded DCO-OFDM, while for RS- OFDM-IM (19) yields two alternative parameters sets {(23,5), (23,6)} with $SE_{RSO}(32,23,5)$=2.06 (bit/s/Hz) and $SE_{RSO}(32,23,6)$=2.09 (bit/s/Hz), respectively. that generate an SE equal or greater than that of coded DCO-OFDM at two message length of $K' = 26$ and $K' = 22$ are calculated. For $K' = 26$, we have $SE_{DCO}(32,26) = 2.03$ (bit/s/Hz) in coded DCO-OFDM, while for RS- OFDM-IM (19) yields two alternative parameters sets {(23,5), (23,6)} with $SE_{RSO}(32,23,5)$=2.06 (bit/s/Hz) and $SE_{RSO}(32,23,6)$=2.09 (bit/s/Hz), respectively. For $K' = 22$ with $SE_{DCO}(32,22) = 1.72$ (bit/s/Hz), RS-OFDM- IM can achieve a similar result using {(19,5), (19,6)} with corresponding SE given by $SE_{RSO}(32,19,5)$=1.75 (bit/s/Hz) and $SE_{RSO}(32,19,6)$=1.78

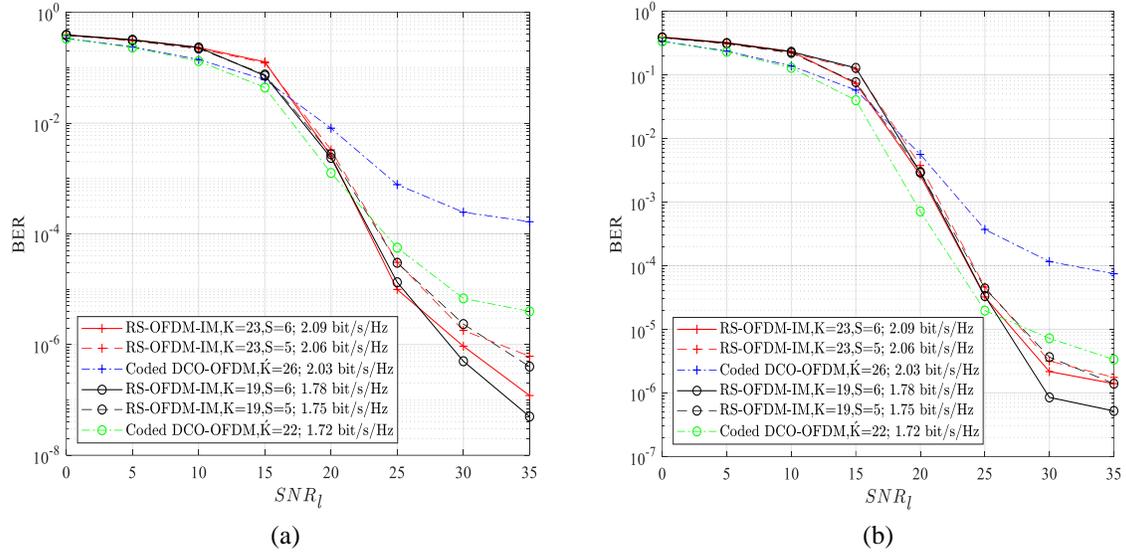

Fig.6 RS Coded DCO-OFDM vs. RS-OFDM-IM in AWGN channel using codewords over GF ($2^5$) and 32-QAM employment ; (a) Double-sided clipping, (b) Single-sided clipping.

(bit/s/Hz). The BER perforamcens of the two scehems with the given parameters under double-sided ($u_{lower} = -2\sigma_x$, $u_{upper} = 2\sigma_x$) and single-sided ($u_{lower} = -1.8\sigma_x$, $u_{upper} \to \infty$) clipping are shown in Fig.6(a) and Fig.6 (b), respectively. It can be seen than RS-OFDM-IM has a weaker performance at low SNR due to the erroneous LLR detection and index demodulation caused by Gaussian noise, however at higher SNR when clipping noise is dominant it gives a significantly improved BER performance compared to coded DCO-OFDM at both SE levels. Note that for coded systems, focus is on the BER values below $10^{-5}$. Fig.6 shows the superiority of proposed scheme in achieving BER $< 10^{-5}$ in both of clipping conditions . For instance, at $SNR_l = 35$ dB in Fig.6 (a), the error floor reduces from $\sim 10^{-4}$ in coded DCO-OFDM with SE=2.03 (bit/s/Hz) to $10^{-7}$ in RS-OFDM-IM with SE= 2.09 (bit/s/Hz) due to the mitigated clipping noise. Fig.6 also shows that, although increasing the puncturing number causes error-correction loss in codewords, it leads to a better overall BER and higher SE in the RS-OFDM-IM. As discussed before, the time-domain signal of RS-OFDM-IM undergoes a less significant distortion due to its reduced envelope when the clipping bounds of LED are adjusted for its illumination purposes. In Section III, we showed that as the dimming level changes, the clipping bounds need to be fixed such that the average optical power of VLC in (16) meets the required LED's dimming level. Given that the optical power in DCO-OFDM scheme is mostly determined by the added biasing signal, when approaches either side of its margin, the clipping bound during bipolar to unipolar conversion on that side will become drastically smaller according to (6) to keep the converted signal within the LED's dynamic range. This will increase clipping noise power and deteriorate the system's BER performance. It was shown in section III that the RS-OFDM-IM will experience less clipping noise due to its reduced time-domain envelope, and therefore a higher SNDR can be obtained when the dimming level varies. To see the impact of this improvement on the overall transmission performance when the parameters of the physical layer are modified for illuminations requirement, the throughput of RS-OFDM-IM and coded DCO-OFDM are evaluated and compared as a function of $\eta$ at two $SNR_t$ levels. Fig. 7 presents the throughput of RS-OFDM-IM given by (30) in an OFDM-based VLC system with optical power limit given by $\Lambda = [0, 5]$, similar to the specification used in Table 1. We consider an LED transmitter in an OFDM system with $N_t$=1024 total subcarriers and 512 independent subcarriers available for data transmission before HS been applied. In the coded DCO-OFDM information are encoded by $RS(31, K', 31-K'+1)$ encoder over $GF(2^5)$, and then codewords are modulated using 32-QAM and transmitted over a Gaussian channel. In RS-OFDM-IM data are encoded using $RS(31, K, 31-K+1)$ encoder where $S$ symbols are punctured according to SIM, and the remaining symbols are modulated by 32-QAM and mapped onto subcarriers. Considering SE of coded DCO-OFDM at two coding rates as the reference, the encoding parameter sets *{(K,S)}* in RS-OFDM-IM are again calculated by (19) to achieve a similar SE. At coding rate $r_c = 83\%$ (K'=26) coded DCO-OFDM yields an SE of 2.03 [bit/s/Hz]. In RS-OFDM-IM two calculated sets $(K_p, S_p)$ = {(24,3), (24,4)} can similarly achieve 2.06 (bit/s/hz) and 2.09 (bit/s/hz), respectively. At $r_c = 70\%$ (K'=22), coded DCO-OFDM yields $SE_{DCO}(22,32) = 1.72$ (bit/s/Hz), while according to (19) the corresponding sets in RS-OFDM-IM will be {(19,5), (20,3)}, where each can achieve an SE of 1.75 (bit/s/Hz). The throughput of the two schemes are measured at two particular $SNR_t$ level since it also accounts for the effect of change in the biasing signal due to the variation in the dimming level. Fig. 7 shows that RS-OFDM-IM supports a higher throughput than coded DCO-OFDM in a broad range of dimming levels. At $SNR_t$=30 dB, RS-OFDM-IM can provide a more consistent throughput during the variation of dimming level. This advantage is due to the fact that RS-OFDM-IM yields a higher SNDR when the dimming level approaches its limits as presented in Fig. 3. It also should be noted that since the $SNR_t$ takes the biasing signal into account, an increase in $I_b$ due to a rise in the illumination level will attenuate the useful electrical signal power and will deteriorate the system BER performance. This effect leads to an asymmetrical form in the throughput curves at $SNR_t = 25$ dB. This attenuation and the residual clipping noise which is more severe at $\eta = 0.9$ and $\eta = 0.1$ lead to erroneous subcarrier index demodulation and subsequently to erroneous RS decoding. This impact will become more pronounced when the puncturing number gets larger due to more loss in the RS code error-correction capability. However, it can be seen that for $0.2 < \eta < 0.9$, RS-OFDM-IM can provide higher throughput at both SNR levels even though the error-correction capability of its block code is compromised by puncturing.

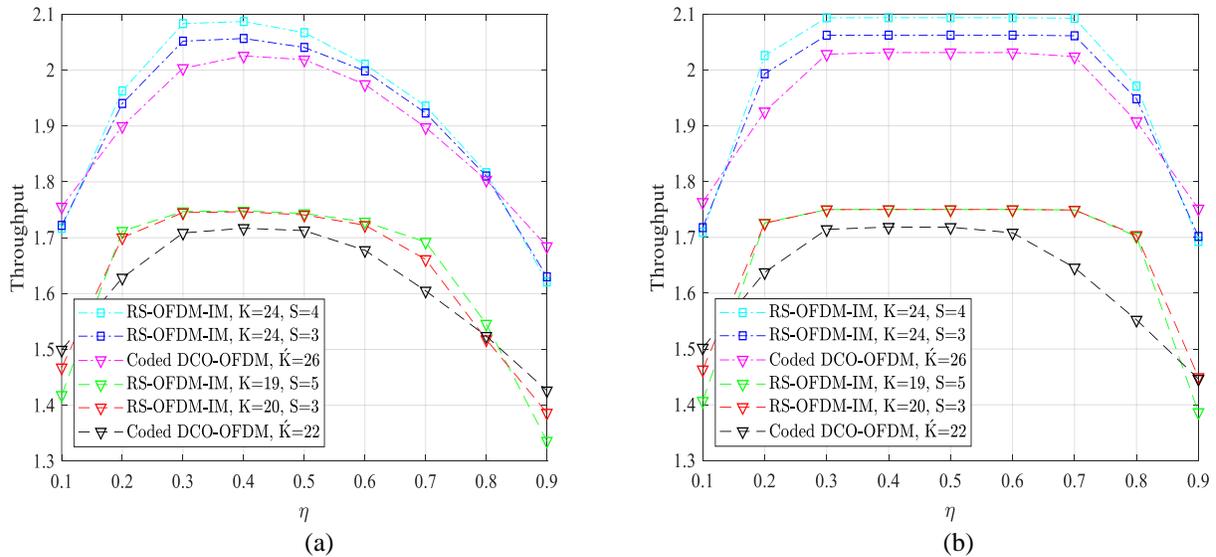

Fig.7 Throughput comparison of RS-OFDM-IM and DCO-OFDM at varying dimming levels; (a) $SNR_t = 25$ dB, (b) $SNR_t = 30$ dB.

## V. CONCLUSIONS

We proposed a novel modulation scheme based on Reed Solomon encoding and subcarrier index modulation for OFDM-based VLC with a physically constrained optical transmitter. In this scheme, the frequency-domain symbols are chosen from the RS encoded block. To overcome the burst error and distortion caused by the nonlinear clipping, a group of codeword symbols are punctured and their corresponding subcarriers are nulled to reduce the clipping probability. The subcarrier index modulation integrated with OFDM is used to compensate for the spectral efficiency loss caused by systematic encoding. The theoretical bound and simulation results show the BER performance superiority of the new scheme in comparison to the traditional OFDM-IM with channel coding and RS coded DCO-OFDM under both single-sided and double-sided clipping conditions.

The parameters involved in the encoding step of the new scheme provides the flexibility for both improving the spectral efficiency toward the desired range and removing the BER floor caused by clipping noise. This feature enables RS-OFDM-IM in achieving and maintaining a better BER and throughput in a physically constrained VLC while the required dimming level is subject to change.